# APPLYING THE AFFECTIVE AWARE PSEUDO ASSOCIATION METHOD TO ENHANCE THE TOP-N RECOMMENDATIONS DISTRIBUTION TO USERS IN GROUP EMOTION RECOMMENDER SYSTEMS

John Kalung Leung[1], Igor Griva[2] and William G. Kennedy[3]

[1]Computational and Data Sciences Department, Computational Sciences and Informatics, College of Science, George Mason University, 4400 University Drive, Fairfax, Virginia 22030, USA
jleung2@gmu.edu

[2]Department of Mathematical Sciences, MS3F2, Exploratory Hall 4114, George Mason University, 4400 University Drive, Fairfax, Virginia 22030, USA
igriva@gmu.edu

[3]Center for Social Complexity, Computational and Data Sciences Department, College of Science, George MasonUniversity, 4400 University Drive, Fairfax, Virginia 22030, USA
wkennedy@gmu.edu

## ABSTRACT

*Recommender Systems are a subclass of information retrieval systems, or more succinctly, a class of information filtering systems that seeks to predict how close is the match of the user's preference to a recommended item. A common approach for making recommendations for a user group is to extend Personalized Recommender Systems' capability. This approach gives the impression that group recommendations are retrofits of the Personalized Recommender Systems. Moreover, such an approach not taken the dynamics of group emotion and individual emotion into the consideration in making top-N recommendations. Recommending items to a group of two or more users has certainly raised unique challenges in group behaviors that influence group decision-making that researchers only partially understand. This study applies the Affective Aware Pseudo Association Method in studying group formation and dynamics in group decision making. The method shows its adaptability to group's moods change when making recommendations.*

## 1. INTRODUCTION

There are two keys design considerations in Group Recommender Systems (GRS). All commercial Recommender Systems in the market are Group Recommender Systems (\cite{pujahari2015group}). Group Recommender Systems would focus its design on making and distributing personalized recommendations to users in the group, and regardless the grouping formation is initiated by users or by the system. Some Group Recommender Systems provide users with group formation functions to self-create, manage, maintain, and disband a group. In most if not all Recommender Systems, for performance and throughput reasons, the system groups individual users or groups of users with similar preferences and tastes into a group without their awareness. When a Group Recommender System makes a top-N recommendation to an active user in a group, the same top-N list will serve all group members. Such a system users' grouping strategy is known as the system simulcast group (SSG) and is the most common type of system users' grouping strategy. Another system users' grouping strategy is a system broadcast group (SBG), which refers to distributing the same information to all SSG. SSG and SBG message distribution is that SSG is intra-group oriented, whereas SBG is inter-group oriented.

The second Group Recommender Systems design point is taking from a user perspective about user grouping types. A user can be a single member of a uni-group (UG) or a multigroup (MG) member. Thus, the top-N list distribution could involve a network of SSG groups where the user is a member. Making top-N recommendations for an active user is relatively straight forward for a Recommender System. All it requires is to access the user's preference profile to guide the top-N recommendations-making process in evaluating and selecting recommendation candidates. However, making recommendations for a group of users is more challenging than making personalized recommendations. The Recommender System must adopt a group decision-making strategy to guide the top-N recommendations making for all group members. In other words, the Recommender System must consider the group preferences profile and the individual active user's preference profile when making the top-N recommendations for a group. Hence, regardless of whether the preference profile is group-oriented or personal oriented, the profile values are ever-changing as the group member consumed an item.

People love socializing. From a user's perspective, a Recommender system should support group formation management functions for group creation, group deletion, adding members to a group, removing members from a group, listing members in a group, setting group member privileges, notifying members in a group, and member vote accounting. Users should not be burden by the system in preference profile maintenance nor respond to query from the system about the user's preferences. The ideal system should know all about its users' likes and dislikes. It can derive the needed users' preference of information by mining the system's implicit metadata, such as system transactions and log files.

From a system perspective, it needs to find a viable solution to support all the above functions and more. For practical purposes, the minimum number of users in a group is two (2). One must find a means to compare the differences of profiles between a pair of users. This study draws the lesson learned from prior works in (Leung, Griva, and William G. Kennedy 2020a) and (Leung, Griva, and Kennedy 2021) that apply the Affective Aware Pseudo Association Method (AAPAM) to facilitate the affective computing of pairwise object's preference profile for the system and users group related operations. This study intends to address what feasible a Recommender System can do to provide the necessary support for the system's and users' group related operations while addressing the ever-changing users' profile issue.

## 2. RELATED WORK IN GROUP RECOMMENDER SYSTEMS

Two or more persons can form a group—for example, friends who meet every weekend for dinner. A group can temporarily include random people to honor a March of Dimes walk for some worthy causes. Regardless of the formation of a group is either a recurring basis or an ephemeral basis, making a recommendation to a group faces two significant challenges: the semantics of group recommendation and the efficient way to compute for group recommendation (Amer-Yahia et al. 2009).

To solve problems such as cold start, data sparsity, and scalability, many commercial websites that deployed Personalized Recommender Systems have incorporated methods by grouping users with similar preferences to share a list of recommendations for each group and (Masthoff 2015). Alternatively, in making a recommendation for groups, conventional approaches in information retrieval and filtering extend Personalized Recommender Systems (Ricci, Rokach, and Shapira 2011). However, this approach gives the impression that group recommendations are retrofits of their personalized counterparts. Although some researchers argue that recommendations for groups using the extension approach perform well, such Group Recommender Systems are not designed with the proper algorithms to solve problems specific to a group setting (Baltrunas, Makcinskas, and Ricci 2010).

## 2.1. State of Group Recommender Systems

Group Recommender Systems (GRS), despite the lack of notoriety in the field of Recommender Systems (RS), have been around as long as their glamorous sibling, the Personalized Recommender Systems (PRS), which were introduced in the mid-1990s to expedite personalized information retrieval on the Web which based on users' preferences (Adomavicius and Tuzhilin 2005). Later, when smartphones become prevalent, research on Mobile Recommender Systems become a significant thrust (Adomavicius and Tuzhilin 2011). Recommender Systems use different information sources to provide users with predictions and recommendations of items while balancing factors like accuracy, novelty, dispersity, and stability in the process. Collaborative Filtering is the favorite choice of method in recommendation processing, which often couples with other information filtering techniques to form a Hybrid Recommender System for achieving higher performance and better user experience (Adomavicius and Tuzhilin 2005), and (Burke 2002). For example, other popular types of information filtering methods include Content-Based, Knowledge-Based, Demographic-Based, and Social Networking Based Kompan et al. (Kompan 2012), Adomavicius et al. (Adomavicius and Tuzhilin 2005), Kywe et al. (Kywe, Lim, and Zhu 2012), De et al. (De Pessemier, Dooms, and Martens 2014), Bobadilla et al. (Bobadilla et al. 2013), and Burke et al. (Burke 2002). Nonetheless, many of the personalized recommended items are often (or mostly) used by groups rather than by individuals, for example, FlyTrap for music (Crossen, Budzik, and Hammond 2002), Pocket RestaurantFinder for restaurants (McCarthy 2002), e-Tourism for tourism (Garcia, Sebastia, and Onaindia 2011), PolyLens for movies (O'connor et al. 2001), LET'S BROWSE for group web surfing (Lieberman, Van Dyke, and Vivacqua 1998), and YuTV for TV shows (Yu et al. 2006) are all Group Recommenders.

In recent years, various Group Recommender Systems have emerged. Besides building from scratch, these Group Recommenders most augment a Personalized Recommender to become a Group Recommender System through one of the two recommendation strategies. The first group recommendation strategy is "aggregating recommendations" of personalized recommendations into recommendations for the whole group (Cantador and Castells 2012). It is similar to the concept, "aggregated predictions," as denoted by Senot et al. (Senot et al. 2010), are the results of aggregating predictions from an individual user into a group prediction. The second group recommendation strategy is "aggregating preferences" of the users' preference model into a group's preference model (Berkovsky and Freyne 2010). Similarly, "aggregated models," as described by Cantador et al. (Cantador and Castells 2011), refer to aggregate individual user data into group data.

From a user perspective, when a Group Recommender System allows users to create and manage groups, the grouping behavior is explicit (Jameson and Smyth 2007). In contrast, a Group Recommender System derives groups through aggregating recommendations or aggregating preferences or aggregated predictions or aggregated models; such grouping behaviors denote as the implicit grouping (Rashid, Karypis, and Riedl 2008). One of the benefits of grouping users into groups is eradicating the cold start problem that all Recommender Systems face (Park and Chu 2009).

## 2.2. Grouping Types

Since different groups exist, group recommender systems aim to manage the heterogeneity of groups. Boratto et al. (Boratto and Carta 2010) speculated that the formation of a group would affect its model and thus the predictive capability of a Recommender System. Building on the work of Jameson et al. ((Jameson and Smyth 2007), which described the four tasks of a recommender system in detail, Boratto et al. (Boratto and Carta 2010) further extended these four tasks to four different variants of a group:

- Established group: people who share common interests and explicitly choose to join a group;
- Occasional group: people who occasionally meet over some activities;
- Random group: people share a resource on occasion without their explicit consent;
- Auto group: People with shared preferences and automatically identified and grouped to share some scared resources.

### 2.2.1. Established Group

O'Connor et al. (O'connor et al. 2001) described the established groups which they observed in PolyLens, a movie Recommender for group use, have a persistent property that users who joined a group not only shared common interests such as movie watching but also actively participated in group activities such as rating watched movies. Once a user joined an established group, the user tends to stay with the group for a long time. Many Group Recommender Systems, such as PolyLens, apply the Collaborative Filtering method to construct user profiles for an individual member in the group and then build the group profile by merging individual profiles from the group members (Kim et al. 2010). To produce recommendations for each group member, PolyLens uses a Collaborative Filtering algorithm to compute each movie's rating score that meets the user's preferences. Movies with the highest recommended rates become the Top-N candidates for recommendation. PolyLens uses a "least misery" selection strategy for making the group's recommendation (O'connor et al. 2001), i.e., the recommended rating for a group is the lowest predicted rating for a movie, to ensure every member is satisfied.

However, some Group Recommenders such as Jukola (O'Hara et al. 2004) and PartyVote (Sprague, Wu, and Tory 2008) are two music recommenders able to make music recommendations to an established social group of people attending a social event. These two music recommenders work without requiring any user profiles. Instead, these recommenders allow any event attendees to express their preferences by selecting a song, album, artist, or genre from a digital music collection. The rest of the group votes to play songs from the selected list. The Recommender computes the probability of the voted songs and plays the song which has the highest probability.

### 2.2.2. Auto Group

Let $v_i$ be the vector of the ratings of user $i$ for the items and $v_j$ be the vector of the ratings of user $j$ for the items. Cosine similarity measures the similarity of $s_{ij}$ between users $i$ and $j$ as expressed in Boratto et al. (Boratto and Carta 2010) proposed group recommendation algorithm that works in four steps:

1) Step 1. Cosine similarity between users' ratings matrixes measures users' similarity. The evaluation procedure (Gfeller, Chappelier, and De Los Rios 2005) illustrates as follows.

$$\psi(u,v) = cos(\vec{u} \odot \vec{v}) = \frac{\vec{u} \odot \vec{v}}{||\vec{u}||_2 \times ||\vec{v}||_2} = \frac{\sum_{i \in I_{u,v}} r_{u,i} r_{v,i}}{\sqrt{\sum_{i \in I_{u,v}} r_{u,i}^2} \sqrt{\sum_{i \in I_{u,v}} r_{v,i}^2}}$$

Where "$\odot$" is denoted as the dot product operator and "$x$" denoted as the multiplication operator.

2) Step 2. Communities detection algorithm proposed by (Blondel et al. 2008) can apply to the user's similarity network and generate partitions of different granularities.

3) Step 3. Rating prediction for items rated by enough users of a group by aggregating the arithmetic mean of users involved for the group. So, for each item $i$, its rating $r_i$ is expressed as:

$$\psi(u,v) = cos(\vec{u} \odot \vec{v}) = \frac{\vec{u} \odot \vec{v}}{||\vec{u}||_2 \times ||\vec{v}||_2} = \frac{\sum_{i \in I_{u,v}} r_{u,i} r_{v,i}}{\sqrt{\sum_{i \in I_{u,v}} r_{u,i}^2} \sqrt{\sum_{i \in I_{u,v}} r_{v,i}^2}}$$

4) Step 4. Ratings prediction for the remaining unrated items by considering both the rating and the similarity ($t_{ij}$) of its top similar items:

$$\psi(u,v) = cos(\vec{u} \odot \vec{v}) = \frac{\vec{u} \odot \vec{v}}{||\vec{u}||_2 \times ||\vec{v}||_2} = \frac{\sum_{i \in I_{u,v}} r_{u,i} r_{v,i}}{\sqrt{\sum_{i \in I_{u,v}} r_{u,i}^2} \sqrt{\sum_{i \in I_{u,v}} r_{v,i}^2}}$$

Where, "$\odot$" is denoted as the dot product operator.

## 2.3. User Perspective in Grouping

From a user's perspective, groups' persistence is related to both usage patterns and privacy issues. If users want to repeatedly receive recommendations for the same group of people, making the group persistent saves time and effort. Moreover, if groups form and dissociate after a single use, the ephemeral approach is better to meet the need. Whether groups are private, known only to group members, or public and accessible to all. Users will behave differently in addressing their privacy, security, net appearance, and perception needs.

### 2.3.1. Pseudo User

Pseudo user is not a new concept in the field of recommender systems. However, the pseudo-user takes on different semantics and roles under different context as described by O'Connor et al. (O'connor et al. 2001), Melville et al. (Melville, Mooney, and Nagarajan 2002), Resnick et al. (Resnick and Varian 1997), Martinez et al. (Martínez et al. 2009), Vozalis et al. (Vozalis and Margaritis 2003), and Lee et al. (Lee, Park, and Park 2008).

### 2.3.2. Pseudo Grouping

In this research study context, pseudo grouping and virtual grouping are synonyms. Research studies use pseudo users in recommender systems, as in Martinez et al. (Martínez et al. 2009). However, no study has investigated leverages pseudo groups as a grouping strategy and plan to apply pseudo grouping to solve the cold start problem.

### 2.3.3. Combine Similar Groups to form Super Group

Several studies of recommender systems have applied Supergroup mostly in the study of the semantic context under content filtering, such as Hu et al. (Hu, Rai12, and Carin 2016). The supergroup concept is yet widely applied in the grouping strategy of recommender systems research.

## 2.4. System Perspective in Grouping

There are times when a Recommender System is the initiator of the grouping strategy. For example, to improve the performance of making recommendations to users, Recommender Systems may select to partition users with similar taste into a group and make the same set of recommendations to all the users in that group to minimize search through the entire rating database or comparing the similarity of pair-users' profiles. In doing so, the Recommender System sacrifices accuracy for performance gain by making similar recommendations to a grouped of individual users.

## 2.5. Ephemeral and Persistence

The notion of ephemeral as a nature of group had been examined exhaustively by O'Connor et al. (O'connor et al. 2001), Jameson (Jameson 2004), Schafer et al. (Schafer, Konstan, and Riedl 1999), and Good et al. (Good et al. 1999). On the other hand, Gartrell et al. (Gartrell et al. 2010), O'Connor et al. (O'connor et al. 2001), Erickson (Erickson 2003), Smeaton et al. (Smeaton and Callan 2005) had researched the persistence and related issues of groups.

## 2.6. Merging Strategies

O'Connor et al. (O'connor et al. 2001) developed PolyLens, a collaborative filtering recommender system designed to recommend movies for a group of viewers rather than individuals. PolyLens merges the individual user's profile among similar taste users to form the group profile and applies the similarity function against the rating matrix for suitable items for making a recommendation to the group. This dissertation research will evaluate other merging strategies besides the simple aggregation approach.

### 2.6.1. Weight Settings for Influential Members

Kim et al. (Kim and Srivastava 2007) articulated that social influence impacts e-Commerce decision-making. Few studies have considered a social influence in an e-Commerce decision support system because until recently, data about social interaction does not adequately capture in e-Commerce. It becomes apparent that the customer decision process is influenced by information from trusted people, not from product manufacturers or recommendation systems. The social influence from high-quality reviews written by previous consumers can have a direct, positive effect on potential consumers' decision making, and this effect can propagate through a social network Huang et al. (Huang and Benyoucef 2013). One of the aims of this dissertation research is to examine a balanced method in extracting the degree of influence by weighing that members exert among each other through decision-making.

## 2.7. Group Recommendation Making Strategies

There are several ways of extending a Personalized Recommender System to a Group Recommender System as described by De et al. (De Pessemier, Dooms, and Martens 2012). Merging strategy and Virtual User strategy are two conventional approaches for grouping. According to Kagita et al. (Kagita, Padmanabhan, and Pujari 2013), there are three ways to implement the Merging strategy: merged profiles, merging recommendation, and merging score. Incidentally, Kagita et al. (Kagita, Pujari, and Padmanabhan 2013) also illuminated the Virtual User strategy. The aspect of grouping strategies for a system is different from that of a user. Next, to examine are the differences between them. Specifically, an examination on the nature of ephemeral group versus persistence group from a user perspective and the following grouping strategies from a system perspective:

1) Making group recommendations for a clustered of users,
2) Making group recommendations by considering transitive precedence relation through a virtual user model,
3) Adapting personalized recommendation for a group through aggregation strategies,
4) Automatic identification groups of users with similar interests, and
5) Making group recommendations through group behavior modeling.

**2.7.1. Making Group Recommendations for a Clustered of Users**

To avoid an exhaustive search through the entire user preference database to match a particular user's preferences, Ntoutsi et al. (Ntoutsi et al. 2012) advocated a group recommendation system that enhances recommendations partitioning users into clusters based on similar preferences. Aggregated preferences of clustered members drive the decision making of recommendations for users. Popescu et al. also studied aggregated user preference (Popescu and Exam 2011) and Masthoff et al. (Masthoff 2015). The algorithms to estimate the relevance of an item for a user deployed by Ntoutsi et al. (Ntoutsi et al. 2012) for the group recommendations framework come in two flavors:

1) In studies of Amer et al. (Amer-Yahia et al. 2009) and Konstan et al. (Konstan et al. 1997). both have illustrated personalized recommendations produced from evaluating relevance scores for unrated items of an unusually active user based on collaborative filtering techniques. However, users typically rate only a few items against the vast amount of the available items. Thus, the notion of support measures the percentage of the active user's friends who have expressed preferences for the item to minimize the rating matrix's skewness. The relevance and support scores are then combined to estimate a recommendation worthiness score of an item for a user.

2) In addition to personalized recommendations, group recommendations deploy a model built on context information of users in the group as illustrated in Amer et al. (Amer-Yahia et al. 2009) by combining all individual users' preferences. Moreover, Ntoutsi et al. (Ntoutsi et al. 2012) further refined these group preferences by three aggregation design methods that carry different semantics. Firstly, the Least Misery design will capture cases where strong user preferences act as a veto. Next, the Fair design will capture more common cases where most of the group members are satisfied. Lastly, the Most Optimistic design will capture cases where the most satisfied member of the group acts as the most influential member. After applying these three design methods appropriately, a Top-K list then uses in making recommendations to users in the group.

**2.7.2. Making Group Recommendations through Aggregation Functions**

Two dominant strategies for making a Personalized Recommender to become a Group Recommender (Berkovsky and Freyne 2010). The first grouping strategy is to aggregate individual preferences into a recommendation list. In effect, this approach creates a pseudo-user for a group based on its group members and then makes recommendations based on the pseudo user's preferences. The second grouping strategy is to aggregate the individual member's recommendation list to form the group recommendation list. In other words, every single user in the group will receive an individual recommendation list. All individual recommendation lists combine to form a recommendation list for the group. The aim of making a group recommendation is to compute a recommendation score for each candidate item that reflects the interests and preferences of all group members. An acceptable approach to obtain a consensus of group ranking recommendation score for a candidate item is through some aggregation functions. Popular aggregation functions are as follows.

Least-misery:

$$\psi(u,v) = cos(\vec{u} \odot \vec{v}) = \frac{\vec{u} \odot \vec{v}}{||\vec{u}||_2 \times ||\vec{v}||_2} = \frac{\sum_{i \in I_{u,v}} r_{u,i} r_{v,i}}{\sqrt{\sum_{i \in I_{u,v}} r_{u,i}^2} \sqrt{\sum_{i \in I_{u,v}} r_{v,i}^2}}$$

Average:

$$\psi(u,v) = cos(\vec{u} \odot \vec{v}) = \frac{\vec{u} \odot \vec{v}}{||\vec{u}||_2 \times ||\vec{v}||_2} = \frac{\sum_{i \in I_{u,v}} r_{u,i} r_{v,i}}{\sqrt{\sum_{i \in I_{u,v}} r_{u,i}^2} \sqrt{\sum_{i \in I_{u,v}} r_{v,i}^2}}$$

Average without Misery:

$$\psi(u,v) = cos(\vec{u} \odot \vec{v}) = \frac{\vec{u} \odot \vec{v}}{||\vec{u}||_2 \times ||\vec{v}||_2} = \frac{\sum_{i \in I_{u,v}} r_{u,i} r_{v,i}}{\sqrt{\sum_{i \in I_{u,v}} r_{u,i}^2} \sqrt{\sum_{i \in I_{u,v}} r_{v,i}^2}}$$

where

$$\psi(u,v) = cos(\vec{u} \odot \vec{v}) = \frac{\vec{u} \odot \vec{v}}{||\vec{u}||_2 \times ||\vec{v}||_2} = \frac{\sum_{i \in I_{u,v}} r_{u,i} r_{v,i}}{\sqrt{\sum_{i \in I_{u,v}} r_{u,i}^2} \sqrt{\sum_{i \in I_{u,v}} r_{v,i}^2}}$$

### 2.7.3. Adapting Personalized Recommendation for Group through Aggregation Strategies

The main problem of group recommendation stems from adopting personalized recommendation based on individual user preference. Recommender Systems have to learn users' preferences from user rating records. After learning individual user preferences, recommender systems then aggregate user rating information to provide group recommendations. Aggregation strategies research is ongoing. Researchers such as Mastho et al. (Masthoff 2015) developed various aggregation strategies that later research built-on. Highlight below is a few of their aggregation strategies. Merging individual user profiles as the group's preferences is the superior strategy.

The utilitarian strategy considers utility values for group recommendation. The utility values are of two types, i.e., additive or multiplicative. An example of a movie recommender system first adds/multiply all the movie ratings separately for a user group. Then display the movies list in the highest aggregate utility value as recommended items.

Most pleasure strategy generates a group rating list based on the maximum of individual ratings. For example, movies with the highest universal rating values will add to the recommended list.

The least misery strategy generates a group rating list based on the minimum of individual ratings. Then the item with a high minimum individual rating will be recommended. The idea behind this strategy is that a group is as happy as its least happy member.

Padmanabhan et al. (Padmanabhan, Seemala, and Bhukya 2011) implemented a rule-based aggregation strategy called RTL strategy to group recommender systems by combining all of the above three strategies. This algorithm performs better regarding accuracy in group recommendation as compared to the preceding three grouping strategies. However, RTL has no provision for creating a better group (regrouping). Pujahari et al. (Pujahari and Padmanabhan 2015) solved the regrouping deficiency by applying Predictive Rule Mining algorithms (Machado 2003), which featured learning rules to learn individual users' preferences applying a new algorithm for making group recommendation.

## 3. DATASETS

Developing an Emotion Aware Recommender System requires an emotion labeled dataset to work with various machine learning algorithms. However, there is no readily available emotion labeled dataset in any publicly accessible repository. One can overcome the deficiency by mining emotion features from existence metadata in the dataset. One can demonstrate making top-N movie recommendations to users in a group through an Emotion Aware Recommender System. Thus, it is necessary to mine the needed emotion label from the subjective text in movie branch mark datasets from the MovieLens repository and The Movie Database (TMDb) website, where the movie overview contains the subjective text metadata. One can download a variety of MovieLens movie datasets stored in the GroupLens repository. However, one needs to scrape the TMDb website for needed metadata. Before performing any emotion modeling on a dataset, one needs to develop a text-based emotion classifier to classify the movie overview's emotion features.

### 3.1. MovieLens Datasets

In the article Leung et al. (Leung, Griva, and William G. Kennedy 2020b), it worked with (Riedl and Konstan 2015) advocated the four MovieLens movie benchmark datasets, namely, ml-latest-small (a.k.a. mlsm), ml-20m (a.k.a. ml20m), ml-25m (a.k.a. ml25m), and ml-latest (a.k.a. ml27m hereafter) datasets. (Leung, Griva, and William G. Kennedy 2020b) also illustrated the need to scrape the TMDb for movie overview metadata to use as input to the Tweets Affective Classifier (TAC) to classify the scraped TMDb dataset's emotion labels before merging with MovieLens datasets. Table 1 illustrates the statistics of each mentioned MovieLens dataset after merging with the emotion labeled TMDb dataset. The name of a MovieLens dataset reflects the number of ratings it contains. Please note that the Number of Users column does not match up the Number of Overviews column. The reason is that the collection of movies in MovieLens and TMDb is not identical.

Table 1. Statistics of MovieLens datasets.

| Attribute Dataset | No. Users | No. Movies | No. Ratings | No. Overviews |
|---|---|---|---|---|
| mlsm | 610 | 9742 | 100836 | 9625 |
| ml20m | 138493 | 27278 | 20000263 | 26603 |
| ml25m | 162541 | 62423 | 25000095 | 60494 |
| ml27m | 283228 | 58098 | 27753444 | 56314 |

In each of the four MovieLens datasets, there are two data files named ratings, and tags contain user ids as a unique identifier. MovieLens stated that user ids found in ratings and tags data files are consistent within the same dataset but are not uniform across different datasets (grouplens.org 2019). For example, in (Leung, Griva, and William G. Kennedy 2020b), user id 400 is only compatible within the same MovieLens MLSM dataset and is not across ml20m, ml25m, ml27m datasets. In other words, user id 400 in other MovieLens datasets are not the same user id 400 as in the MLSM dataset. However, (Leung, Griva, and William G. Kennedy 2020b) has demonstrated by using the Affective Aware Pseudo Association Method (AAPAM), the disjoint user id 400 in MLSM can correctly connect to the proper user id in MovieLens datasets, as depicted in Table 2. For example, user id 400 of MLSM can Pseudo Associate Connect (PAC) to user id 66274 of MLSM or PAC to user id 95450 of ml25m or PAC to user id 89195 of ml27m. All mentioned user id are disjoint users of each other. In the context of this study, disjoint users refer to two or more individual mutually exclusive users whose user id are different and reside in different datasets within the same or different information domains whose emotion profiles, UVEC, are highly similar or even identical. One can consider the disjoint users as identical pseudo users.

### 3.2. Affective Aware Pseudo Association Method

The Affective Aware Pseudo Association Method (AAPAM) computes the Affective Index Indicator (AII) using the Cosine Similarity algorithm (Bigdeli and Bahmani 2008), as depicted in Equation 2, Cosine Similarity expresses the closeness of the emotion profiles between two users or items. When using AAPAM to compare pairwise between User id 400 of MLSM against users in other MovieLens datasets, AII reveals, as depicted in Table 2, the closest other users' emotion profiles that match the candidate user. User id 400 in MLSM can make a pseudo associate connection (PAC) to user id 66274 with AII 0.999916 in ml20m or to user id 95449 with AII 0.999999 ml25m, or user id 89195 with AII 0.999999 in ml27m, respectively.

$$Inner(x, y) = \sum_i x_i y_i = <x, y> \quad (1)$$

$$CosSim(x, y) = \frac{\sum_i x_i y_i}{\sqrt{\sum_i x_i^2} \sqrt{\sum_i y_i^2}} = \frac{<x, y>}{||x|| \, ||y||} \quad (2)$$

AAPAM also worked with The Movie Database (TMDb) (TMDb 2018). By scraping from TMDb, it yields the movie metadata for movie overviews, poster images, and other metadata. AAPAM applied the Tweets Affective Classifier (TAC), a method developed in (Leung, Griva, and William G Kennedy 2020), to classify a movie emotion profile. A movie emotion profile is also known as a movie vector, MVEC, which represents a multi-dimensional embedding of a probability distribution of seven primary human emotions: neutral, happiness, sadness, hate, anger, disgust, and surprise. Each user in (Leung, Griva, and William G Kennedy 2020) also has a user emotion profile, UVEC, where it contains the average value of all movies MVECs the user has watched.

Table 2. Pseudo Association Connection of ml-latest-small user id 400 to other users in different datasets through affective index indicator.

| **Dataset** | **mlsm** | **ml20m** | **ml25m** | **ml27m** |
|---|---|---|---|---|
| User1 ID PAC | 400 | 66274 | 95459 | 89195 |
| User1 Movie Count | 43 | 22 | 43 | 43 |
| User1 Watched List movieID | 6<br>47<br>50<br>260<br>…,<br>122886<br>134130<br>164179<br>168252 | 47<br>260<br>300<br>307<br>...,<br>2628<br>2797<br>3418<br>3481 | 6<br>47<br>50<br>260<br>…,<br>122886<br>134130<br>164179<br>168252 | 6<br>47<br>50<br>260<br>…,<br>122886<br>134130<br>164179<br>168252 |
| User1 UVEC<br>Neutral<br>Happiness<br>Sadness<br>Hate<br>Anger<br>Disgust<br>Surprise | 0.16353<br>0.08874<br>0.12709<br>0.20332<br>0.11934<br>0.15881<br>0.13918 | 0.16250<br>0.08609<br>0.12654<br>0.20701<br>0.11776<br>0.16005<br>0.14005 | 0.16353<br>0.08874<br>0.12709<br>0.20332<br>0.11934<br>0.15881<br>0.13918 | 0.16353<br>0.08874<br>0.12709<br>0.20332<br>0.11934<br>0.15881<br>0.13918 |
| User1 Affective | 1.0 | 0.99992 | 0.99999 | 0.99999 |

| Index Indicator | | | | |
|---|---|---|---|---|

As illustrated in Table 2, user id 400 in the rating data file of the MLSM dataset has watched 43 movies; taking the average of all the 43 movies' MVECs yields the UVEC for user id 400. As mentioned in (Leung, Griva, and William G Kennedy 2020), a movie's MVEC is static and stays unchanged throughout the film's life; whereas, a user's UVEC changes its value each time the user watches a movie. The user's UVEC reflects the up-to-date movie taste and preference of the user. A movie MVEC is unique, while a UVEC may not be unique when two users watched the same movie set.

Besides, the AAPAM method can PAC connect disjoint users from different datasets within the same domain; this study believes the same technique can PAC connect disjoint users and items among different datasets across different domains. Unlike MovieLens datasets, some other movie datasets such as TMDb highlight the average voting score, the sentiment rating value on a scale of 1 (lowest) to 10 (highest), of a movie by a group of users who have watched and rated the movie through the voting count attribute in the data file instead of individual user's sentiment. TMDb does not maintain user information, and neither contains a user-id field in the dataset. When applying the AAPAM to connect MovieLens and TMDb domains, the PAC connection applies to movie items between MovieLens and TMDb. Here, the PAC connection between movie A in MovieLens to movie B in TMDb indicates how similar the two movies' emotion profiles, MVECs, form a one-to-one relationship. However, when applying the PAC to connect a user in MovieLens and a movie item in TMDb, the movie item MVEC in TMDb must first be normalized with the respective voting count. The normalized MVEC represents the average UVEC of the group of users who have rated it. Thus, the PAC connection between user A in MovieLens to the normalized MVEC of movie B in TMDb indicates how similar user A to a group of users B is in the form of a one-to-many relationship.

## 4. METHODOLOGY

This study utilized affective features in two data sources that it uses. The affective features applied to users' emotion profiles, UVECs, and items' emotion profiles, IVECs, or MVECs equivalent. No disjoint users and items can interconnect without adding affective features across data sources of MoieLens, and TMDb. Affective aware features are added to the data sources through Tweets Affective Classifier (TAC) as developed in (Leung, Griva, and William G Kennedy 2020). Table 3 depicts samples after added affective features UVEC and MVEC in this study's data sources.

Table 3. TMDb movie emotion profile example.

| tmdbId | 2 | 525662 |
|---|---|---|
| movieId | 4470 | 189111 |
| Mood | Disgust | Hate |
| Neutral | 0.15705037 | 0.11876434 |
| Happiness | 0.08608995 | 0.05086204 |
| Sadness | 0.15583897 | 0.12669845 |
| Hate | 0.07506061 | 0.3391073 |
| Anger | 0.08469571 | 0.13069303 |
| Disgust | 0.26612538 | 0.13746719 |
| Surprise | 0.17513901 | 0.096407644 |

### 4.1. Tweets Affective Classifier

Leung et al. in (Leung, Griva, and William G Kennedy 2020) illustrated the Tweets Affective Classifier (TAC) development, a text-based tweets emotion classifier capable of detecting and recognizing six basic human emotions advocated by Paul Ekman. The six emotions are happiness, sadness, fear, anger, disgust, and surprise. For affective computing convenience, one designs TAC capable of detecting and recognizing natural emotion. One then feeds TMDb's movie overview to TAC to classify the overview's moods or the movie's emotion profile, MVEC depicted in Table 4. After joining the emotion-labeled TMDb data file with the rating data file in MovieLens, it yields an emotion-labeled MovieLens data file depicted in Table 5. With the joined emotion-labeled data file, one computes the users' emotion profiles, UVEC, by taking the MVEC from all the movies that the user has watched.

Table 4. First few rows of the cleansed emotion labeled MovieLens data file.

| Index | tid | mid | iid | mood | neutral | happy | sad | hate | anger | disgust | surprise |
|---|---|---|---|---|---|---|---|---|---|---|---|
| 1 | 2 | 4470 | 94675 | disgust | 0.157 | 0.086 | 0.156 | 0.075 | 0.085 | 0.266 | 0.175 |
| 2 | 5 | 18 | 113101 | disgust | 0.121 | 0.060 | 0.098 | 0.128 | 0.133 | 0.244 | 0.216 |
| 3 | 6 | 479 | 107286 | hate | 0.075 | 0.114 | 0.054 | 0.433 | 0.095 | 0.128 | 0.100 |
| 4 | 11 | 260 | 76759 | neutral | 0.299 | 0.262 | 0.079 | 0.030 | 0.017 | 0.083 | 0.230 |
| 5 | 12 | 6377 | 266543 | surprise | 0.150 | 0.080 | 0.055 | 0.083 | 0.103 | 0.153 | 0.376 |

Table 5. Sample if users' emotion profiles derived from taking the average of the user's moods values from all movies the user has watched

| mlsm | neutral | joy | sadness | hate | anger | disgust | surprise |
|---|---|---|---|---|---|---|---|
| 400 | 0.163529 | 0.088735 | 0.1270899 | 0.203318 | 0.119338 | 0.158812 | 0.1391753 |
| 414 | 0.166351 | 0.097305 | 0.1180924 | 0.164195 | 0.115177 | 0.172503 | 0.1663736 |
| 474 | 0.168858 | 0.099746 | 0.1187206 | 0.160877 | 0.112612 | 0.171919 | 0.1672649 |
| 448 | 0.172833 | 0.096858 | 0.1160457 | 0.161207 | 0.112276 | 0.170985 | 0.1697938 |

### 4.2. System Grouping

This paper advocates a method for the system to group users in a system simulcast group (SSG) according to the similarity in users' emotion profiles, UVEC, so that all SSG users share the same top-N recommendations list. However, to simulate the personalized recommendation functionality before pushing the SSG top-N list to an active user in the SSG, the system reranked the top-N list's MVEC against the active user's UVEC, thereby ensuring all users in the SSG receives personalized individual top-N recommendations list. The benefit of grouping users in SSG is improving the system's throughput while performing fewer top-N recommendations made by the Recommender System. For example, a system hosts $n$ million users. The system must run a top-N recommendation process per-user or n million top-N recommendations operations to make $O(n)$. personalized recommendations. By grouping users into $m$ users per SSG, the system only needs to make $\frac{n}{m}$ top-N or $O\left(\frac{n}{m}\right)$. It is an $m$ fold decrease in top-N recommendations making. Thus, by grouping users in SSG, the system can expand the handling of $\approx m$ more users without upsizing the system.

This paper has devised the following scheme to demonstrate an SSG's working in an Emotion Aware Recommender System. MovieLens ml-latest-small (MLSM) ratings data file contains 100,836 ratings of 9743 movies by 610 users. Some users rated 20 movies, the minimum number of movies a user must have rated in MovieLens sampled datasets, while some users have rated thousands of movies. By sorting the MLSM rating data file in descending order with users rated most movies at the top, draw out the top ten users to act as the dominant user in one of the ten

SSG groups. Table 7 depicts ten dominant users to act as anchor user for the ten SSG group. Table 7 also shows the sample list of SSG group members. With each SSG's dominant user's UVEC, it computed against all undraw users through pairwise Cosine Similarity of UVEC. It ranked the result in descending order. In effect, the result shows the Affective Index Indicator (AII) of each member's UVEC relative to the UVEC of the dominant user. Draw out the top 60 users who have the most similar emotion profile of the dominant user to join the SSG. Continue the same step to draw out members for the other nine SSG. The grouping scheme forming ten SSG each has a dominant user who rated most movies and 60 members with closely matched emotion profiles against the dominant user's UVEC.

IMDb periodically publishes a list containing the "IMDb 100 Greatest Movies of All Time." Using the list as top-N recommendations, one can feed to all ten SSG. Doing so, one simulates the system broadcast group (SBG) broadcast message function. Before sending the top-N recommendations to an SSG member, the system reranked the user's UVEC against all movie emotion profiles, MVEC, on the list through the pairwise Cosine Similarity. This paper picks three users from one of the SSG to demonstrate the reranking process. Hence the SSG organized by ranking all members' UVEC by computing their Affective Index Indicator (AII). The dominant member occupies the top of the AII list. The least-misery member occupies the bottom of the AII list, and the 30th member in the AII list is the average member of the SSG.

Table 6 showing the reranked top 10 of the "IMDb 100 Greatest Movies of All Time" for MovieLens ml-latest-small user ID 414 using UVEC of 20% and 100% movie watching history to compute the pairwise AII of Cosine Similarity. The two reranked lists of user ID 414 show a different rank order from the movie ranking of IMDb and the more movies the user has watched. Users with different UVEC affect the reranked result.

Table 6. User ID 414 Reranked IMDb 100 Greatest Movies of All Time using 20% and 100% Movie Watching UVEC

| Ranking 100 All Time Greatest Movies | Moive ID | Movie Title | UserId414 Reranked (UVEC 20% watched) | UserId414 Reranked Movie Title (UVEC 20% Watched) | UserId414 Reranked (UVEC 100% watched) | UserId414 Reranked (AII - UVEC 100% Watched) |
|---|---|---|---|---|---|---|
| 1 | 858 | The Godfather (1972) | 1252 | Chinatown (1974) | 1252 | Chinatown (1974) |
| 2 | 1221 | The Godfather: Part II (1974) | 1213 | Goodfellas (1990) | 1213 | Goodfellas (1990) |
| 3 | 2019 | Seven Samurai (1954) | 899 | Singin' in the Rain (1952) | 318 | The Shawshank Redemption (1994) |
| 4 | 296 | Pulp Fiction (1994) | 318 | The Shawshank Redemption (1994) | 899 | Singin' in the Rain (1952) |
| 5 | 1203 | 12 Angry Men (1957) | 912 | Casablanca (1942) | 8125 | Sunrise (1927) |
| 6 | 5618 | Spirited Away (2001) | 8125 | Sunrise (1927) | 26150 | Andrei Rublev (1966) |
| 7 | 527 | Schindler's List (1993) | 26150 | Andrei Rublev (1966) | 912 | Casablanca (1942) |
| 8 | 912 | Casablanca (1942) | 89759 | A Separation (2011) | 89759 | A Separation (2011) |
| 9 | 1219 | Psycho (1960) | 1251 | 8½ (1963) | 1237 | The Seventh Seal (1957) |
| 10 | 1213 | Goodfellas (1990) | 1237 | The Seventh Seal (1957) | 1251 | 8½ (1963) |

### 4.3. User Grouping

One significant difference between a Group Recommender System and Personalized Recommender System is that Group Recommender System supports users with functions to initiate group formation, and Personalized Recommender System cannot A Group Recommender System allows the system and users to create a multi-group (MG) and explicitly join the group. The system provides functions for the user to self manages group administration and group maintenance. An MG's joining can be by invitation only known as private multi-group (PMG) or open to public multi-group (OMG). When the system makes a top-N recommendation to an MG, all MG members share the top-N list, and no need to re-ranked the top-N is necessary for individual users in the MG. However, the Group Recommender System may apply a decision-making support strategy in top-N recommendations to assist the group decision making in selecting a recommendation in the top-N list. This study focuses on applying affective awareness recommendations to support group decision-making by the most dominant member strategy, average members' mood strategy, and the Least-misery member strategy. The most dominant member in an MG has the largest number of movies a member has watched among the group. The least-misery member is the member whose emotion profile is most dissimilar to the MG's most dominant member. The least-misery member of the group determines by the pairwise Affective Index Indicator (AII) of all members' emotion profiles in the MG against the most dominant member emotion profile, then rank the AII list in the descending order, the least-misery member occupies the lowest rank. Depicts in Table 8 is a user MG of size 5. User ID 195 is the dominant member of the group, and user ID 463 is the least-misery member.

## 5. Results

### 5.1. Group Formation by System

One needs to determine the number of simulcast group (SSG), g, that wants to form. The system can make a rank list on user interaction and pick a reasonable number for *g*. In the MovieLens dataset, the system can compute the number of movies watching history as the system interaction criteria. Each SSG will anchor with a member pick from the top of the interaction rank list. The system computes the pairwise Cosine Similarity between the anchor member of the SSG against other ungrouped users. After ranking the pairwise list, the system will move *m* number of users from the top of the ranked list to the SSG. Depicted in Table 7 is 10 SSG on MovieLens ml-latest-small dataset. Each SSG anchors with a dominant user with most movie watching history. Each SSG contains 61 members, including the anchor member. The system picks members by the Affective Index Indicator (AII) values highly similar to the anchor member.

Table 7. 10 System Simulcast Groups Formed by Using MovieLens ml-latest-small

| Rank | Dominant User ID | Movie Count | System Simulcast Group Member User ID | Member Count |
|---|---|---|---|---|
| 1 | 414 | 2698 | (1) 414, (2) 232, …, (31) 212, …, (61) 167 | 61 |
| 2 | 599 | 2478 | (1) 599, (2) 477, …, (31) 198, …, (61) 474 | 61 |
| 3 | 474 | 2108 | (1) 474, (2) 560, …, (31) 260, …, (61) 350 | 61 |
| 4 | 448 | 1864 | (1) 448, (2) 226, …, (31) 453, …, (61) 389 | 61 |
| 5 | 274 | 1346 | (1) 274, (2) 330, …, (31) 424, …, (61) 477 | 61 |
| 6 | 610 | 1302 | (1) 610, (2) 160, …, (31) 405, …, (61) 218 | 61 |
| 7 | 68 | 1259 | (1) 68, (2) 414, …, (31) 212, …, (61) 593 | 61 |
| 8 | 380 | 1218 | (1) 380, (2) 160, …, (31) 218, …, (61) 318 | 61 |
| 9 | 606 | 1115 | (1) 606, (2) 177, …, (31) 6, …, (61) 66 | 61 |
| 10 | 288 | 1055 | (1) 288, (2) 483, …, (31) 555, …, (61) 167 | 61 |

### 5.2. Grouping by Users

Once a Group Recommender System provides users with group formation functionality, users can use the group formation to create and maintain a multi-users group (MG). In the following illustration, all members subsampled randomly from the derived emotion labeled movie dataset MLSM. Depicted below is a five-member multi-user group. The group average UVEC derived from the mean of five members' UVEC. User ID 195 watched most movies, thus the dominant member, whereas user ID 463 is the least-misery member.

Table 8. A Five Member Multi-User Group

| Rank | UserId | Watched | UVEC |
|---|---|---|---|
| 1 | 195 | 187 | 0.1639455, 0.0902557, 0.1176815, 0.1726736, 0.1185870, 0.1777129, 0.1591437 |
| 2 | 602 | 135 | 0.1639545, 0.0869860, 0.1168919, 0.16947266, 0.1156349, 0.1817310, 0.1653290 |
| 3 | 190 | 66 | 0.1603803, 0.0849701, 0.1254172, 0.17182250, 0.1135154, 0.1797844, 0.1641099 |
| 4 | 521 | 40 | 0.1574143, 0.0944750, 0.1240710, 0.14589457, 0.1083259, 0.1795868, 0.1902323 |
| 5 | 463 | 33 | 0.1558253, 0.0968441, 0.1140474, 0.19975860, 0.1226243, 0.1571110, 0.1537890 |
| Average Group | | UVEC | 0.1603040, 0.0907061, 0.1196220, 0.17192440, 0.1157376, 0.1751852, 0.1665208 |

## 5.3. Group Decision Making Strategies

A reasonable way to handle a group decision making using the dominant member strategy is to generate the top-N recommendations based on the dominant user's preference. Similarly, to support the group decision making using the least-misery member strategy, the top-N recommendations are generated based on the least-misery member's preference. Instead of using a movie database to generate a disparate top-N list based on different decision-making strategies, this paper limits the overall movie selection range to generate top-N listed from the "IMDb 100 Greatest Movies of All Time". Each of the movies in the IMDb list has computed an emotion profile MVEC. By computing the Affective Index Indicator (AII) pairwise between the UVEC of the preferred user group decision-making strategy and the IMDb MVEC and reranked the result in descending order will yield the desire top-N list. Depicted in Table 9 listed the top-10 of dominant user ID 195, least-misery user ID 463, and using the five-member MG average UVEC. Table 9 shows the top-10 recommendations list generated to support MG users' decision-making strategies of the dominant member, least-misery member, and average group user profile.

Table 9. Top-10 Generated by Dominant Member, Least-misery Member and Average Group User Profile Decision-making Strategies

| Top-10 Rank | UserId195 Dominant User Strategy Top-10 | UserId195 Dominant user Top-10 Movie Title | UserId463 Least-misery User Strategy Top-10 | UserId463 Least-misery User Top-10 Movie Title | Average Group User Strategy Top-10 | Average Group User Top-N Movie Title |
|---|---|---|---|---|---|---|
| 1 | 858 | The Godfather (1972) | 1252 | Chinatown (1974) | 1252 | Chinatown (1974) |
| 2 | 1221 | The Godfather: Part II (1974) | 1213 | Goodfellas (1990) | 1213 | Goodfellas (1990) |
| 3 | 2019 | Seven Samurai (1954) | 899 | Singin' in the Rain (1952) | 318 | The Shawshank Redemption (1994) |
| 4 | 296 | Pulp Fiction (1994) | 318 | The Shawshank Redemption (1994) | 899 | Singin' in the Rain (1952) |
| 5 | 1203 | 12 Angry Men (1957) | 912 | Casablanca (1942) | 8125 | Sunrise (1927) |
| 6 | 5618 | Spirited Away (2001) | 8125 | Sunrise (1927) | 26150 | Andrei Rublev (1966) |
| 7 | 527 | Schindler's List (1993) | 26150 | Andrei Rublev (1966) | 912 | Casablanca (1942) |

| 8 | 912 | Casablanca (1942) | 89759 | A Separation (2011) | 89759 | A Separation (2011) |
| 9 | 1219 | Psycho (1960) | 1251 | 8½ (1963) | 1237 | The Seventh Seal (1957) |
| 10 | 1213 | Goodfellas (1990) | 1237 | The Seventh Seal (1957) | 1251 | 8½ (1963) |

## 6. FUTURE WORK

The author et al. plans to widen the study of applying the Affective Aware Pseudo Association Method in other aspects of Group Recommender Systems besides group formation and group decision-making support. The current study relied on pre-processed users' and items' emotion profiles. Future studies will investigate the real-time processing of objects' emotion profile to accommodate the new arrival of objects to the Recommender Systems.

## 7. CONCLUSION

Human preference in decision-making is highly influenced by their moods. By capturing and modeling emotion features in users and items can better reflect human preferences. This paper advocated applying the Affective Index Indication (AII) developed by the Affective Aware Pseudo Association Method (AAPAM) to obtain pairwise similarity metrics for comparing the closeness between two objects. In the study, the AII was applied to compare two users' emotion profiles UVEC to determine the closeness between users. The same technique was also applied to find the closeness between a user's emotion profile UVEC and a list of movies' emotion profiles MVEC. Provided that movies and users have precomputed with MVEC and UVEC, to make movie top-N recommendations by AII technique is straight forward. The AII generated top-N list also capable of adapting to the dynamic of the user's mood change. AII technique can support various decision-making strategies from group users and support system and user-oriented group formation.

**Authors**

John K. Leung is a Ph.D. candidate in Computational and Data Sciences Department, Computational Sciences and Informatics at George Mason University in Fairfax, Virginia. He has over twenty years of working experience in information technology research and development capacity. Formerly, he worked in the T. J. Watson Research Center at IBM Corp. in Hawthorne, New York. John has spent more than a decade working in Greater China, leading technology incubation, transfer, and new business development.

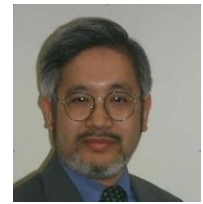

Igor Griva is an Associate Professor in the Department of Mathematical Sciences at George Mason University. His research focuses on the theory and methods of nonlinear optimization and their application to problems in science and engineering.

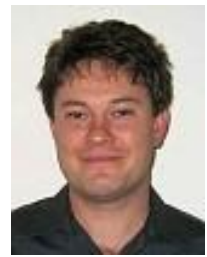

William G. Kennedy, PhD, Captain, USN (Ret.) is an Associate Professor in the Department of Computational and Data Sciences and is a Co-Director of the Center for Social Complexity at George Mason University in Fairfax, Virginia. He has over 10-years' experience in leading research projects in computational social science with characterizing the reaction of the population of a mega-city to a nuclear WMD event being his most recent project. His teaching, research, and publication activities are in modeling cognition and behavior from individuals to societies.

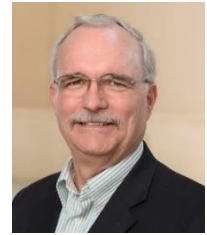